\pdfoutput=1
\newcommand{\UV}{F_{\rm PDC}}  			
\newcommand{\Si}{F_{\rm SFG}}                   
\newcommand{\DeltaPDC}{\Delta^{\rm PDC}}                   
\newcommand{\DeltaSFG}{\Delta^{\rm SFG}}                   
\newcommand{\cV}{{\cal V}_{\rm PDC}^{(\rm inc)}}
\newcommand{\cW}{{\cal W}_{\rm SFG}^{(\rm inc)}}
\newcommand{\vka}{\vec{w}}			
	
\newcommand{\vkap}{\vec{w}\,'}

\newcommand{\im}{i}

\newcommand{\sinc}{{\rm sinc}}
\newcommand{\q}{\vec{q}}

\newcommand{\nn}{\nonumber}
\newcommand{\bsub}{\begin{subequations}}
\newcommand{\esub}{\end{subequations}}
\newcommand{\beq}{\begin{equation}}
\newcommand{\eeq}{\end{equation}}
\newcommand{\beqa}{\begin{eqnarray}}
\newcommand{\eeqa}{\end{eqnarray}}
\newcommand{\beql}{\begin{subequations}\begin{eqnarray}}
\newcommand{\eeql}{\end{eqnarray}\end{subequations}}
\documentclass[pra,onecolumn,aps,amsmath,nofootinbib]{revtex4}
\usepackage{amsmath}
\usepackage{graphicx}
\usepackage{float}   
\usepackage{verbatim}  
\usepackage{braket}
\begin{document}
\title{Space-time coupling in the up-conversion of broadband down-converted light}
\author{E.~Brambilla$^1$,  O. Jedrkiewicz$^{2}$, P. Di Trapani$^{1}$  and A.~Gatti$^{2,1}$}
\affiliation{$^1$ Dipartimento di Scienza e Alta Tecnologia\text{,} Universit\`a dell'Insubria, Via Valleggio 11 22100 Como, Italy, \\
$^2$ CNR, Istituto di Fotonica e Nanotecnologie, Piazza Leonardo da Vinci 32, 20133 Milano, Italy}
\begin{abstract}
We investigate the up-conversion process of broadband light from parametric down-conversion (PDC), focusing on the spatio-temporal spectral properties of the 
sum-frequency generated (SFG) radiation.
We demonstrate that the incoherent component of the SFG spectrum is characterized by a skewed cigar-shaped geometry in space-time 
due to the compensation of spatial walk-off and group-velocity mismatch between the fundamental and the generated SFG fields.
The results are illustrated both by a theoretical modeling of the optical system and by experimental measurements.
\end{abstract}
%
\centerline{Version \today}
\maketitle

\section*{Introduction}
\label{sec:intro}
The process of sum frequency generation occurring in a $\chi^{(2)}$ crystal has often been used in ultra-fast nonlinear optics experiments for probing the spatio-temporal structure of femtosecond pulses. In particular, the three-dimensional mapping of ultrashort complex
pulses \cite{minardi2004,trull2004} and the observation of the spatio-temporal dynamics of Kerr media filamentation \cite{majus2010} has been obtained with this method.

Recent works \cite{dayan2005,peer2005b,ribeiro2006,harris2007,dayan2007} have shown
that sum-frequency generation represents also a resource for exploring and manipulating the entanglement properties of a broadband PDC source,
as well as for applications in quantum information processing.
For example, spectrally engineered SFG has been proposed as a tool for selecting broadband quantum modes (Schmidt modes)
from the ultra-fast multi-mode state generated by PPLN waveguides \cite{silberhorn2011}.

Recently our research group applied the SFG technique for investigating the whole spatio-temporal structure of the PDC biphoton correlation 
\cite{brambilla2012,jedr2012a,jedr2012b} in order to reveal its nontrivial geometry in the space-time domain studied in \cite{gatti2009,caspani2010,brambilla2010}. In that experiment,
a quasi-stationary pump pulse with a broad transverse cross-section was injected in a first $\chi^{(2)}$ crystal and the
generated PDC light was imaged into an identical crystal tuned for the same phase-matching conditions where the SFG process takes place. A perfect tuning between the two crystals allows the partial reconstruction of the original pump beam through the inverse PDC process, i.e. SFG from conjugate mode pairs \cite{jedr2011}.
In such a detection scheme, the SFG crystal is used as an ultra-fast optical correlator and we showed that the coherent component of the SFG radiation
is able to provide detailed information on the biphoton correlation function characterizing the PDC source.

In this work we shall focus on the spatio-temporal properties of the incoherent component of the up-converted PDC field.
In addition to the SFG coherent process which leads to the reconstruction of the original pump beam,
the random up-conversion of photon pairs that do not belong to twin modes
generates SFG radiation over a much broader range of spatio-temporal modes.
This incoherent process, though negligible compared to the coherent process as long as the PDC source operates in a coincidence count regime \cite{dayan2005}, becomes relevant at high parametric gains, i.e when the mean number of PDC photons per mode is larger than unity.
We investigate the spatio-temporal properties of the field produced through this incoherent process, showing
how propagation effects inside the SFG crystal rapidly selects the spatio-temporal modes undergoing the up-conversion process in a non-trivial way, leading to a coupling between the spatial and the temporal frequencies.
We shall demonstrate how the SFG incoherent spectrum displays a nontrivial skewed geometry in the spatio-temporal frequency domain deriving from
the interplay of group-velocity mismatch (GVM) and spatial walk-off between the fundamental (PDC) field and the up-converted one (SFG radiation).

In Sec.\ref{sec:exp_scheme} we illustrate the experimental scheme implemented for the measure of the spatio-temporal spectrum of the up-converted PDC field. The theory developed in Sec.\ref{sec:model}, valid within the quasi-stationary plane-wave pump approximation (PWPA), allows
to obtain approximate expression for the coherent and the incoherent spectra and provide a physical insight into  the experimental
observation. The results obtained from a 3D+1 numerical modeling of the optical system are presented in Sec.\ref{sec:spectrum_num}
and interpreted in the light of the PWPA theory. The results of the experiment are reported in Sec.\ref{sec:exp}.

\section{Scheme of the measurement}
\label{sec:exp_scheme}
\begin{figure}[H]
\centering
{\scalebox{0.45}{\includegraphics*{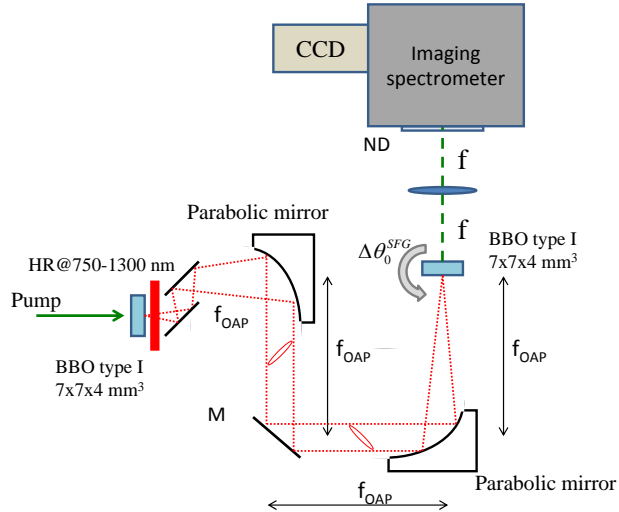}}}
\caption{(color online) Experimental setup for probing the PDC field via sum-frequency generation. The output face of the PDC crystal is imaged at
the input face of a second crystal, identical to the first, where the SFG process takes place. The second crystal
is mounted on a micrometric rotation stage that allows to change the phase-matching conditions in the up-conversion process.
The SFG far-field, obtained through the f-f lens system beyond the second crystal, is analyzed by an imaging spectrometer.}
\label{fig_setup}
\end{figure}

The conditions of the experiment are similar to those reported in \cite{jedr2011}. The PDC radiation is generated in a pulsed regime with high parametric gain $(g\approx 9)$ by a type I BBO crystal pumped at 527.5nm (e-oo phase-matching).  The crystal is cut for collinear emission at the degenerate wavelength $2\lambda_0=1055$nm.
The optical setup is shown in Fig.\ref{fig_setup}.
The pump beam injected in the PDC crystal is a $\sim 1$ps Gaussian pulse with a 500 micrometers transverse beam waist.
The pump is then subtracted with a glass filter (transmission bandwidth 750-1300nm)
and the PDC field is imaged at the entrance face of the SFG crystal where the up-conversion process takes place.
The 4-f imaging device is built with two achromatic parabolic mirrors rather than with lenses, in order to minimize dispersion 
between the two crystals. As demonstrated in \cite{jedr2011,jedr2012a}, it is capable of reconstructing the PDC field
at the SFG crystal input face preserving both its amplitude and its phase over a very broad range of temporal and spatial frequencies.
The temporal and angular bandwidths are respectively on the order of $\Delta \lambda\sim 600$ nm and $\Delta \alpha \sim \pm 4^{\circ}$.

The spatio-temporal features of the output field from the SFG crystal are analyzed in the far-field, by means of
of an imaging spectrometer (IS) that collects the photons 
in the focal plane of a 20 cm focal length lens. A CCD camera (Roger Scientific) is placed at the very output of the spectrometer
(LOT Oriel) in order to record the resulting spectral images.
(see Fig.\ref{fig_setup}).

\section{Theory - spectral characterization of the up-converted PDC light}
\label{sec:model}
We introduce the theoretical framework that allows the interpretation of spectral measurements
obtained from the optical scheme illustrated in the previous section.
A detailed derivation of the PWPA results illustrated in the next subsection can be found in \cite{brambilla2012}.

The results that follow hold in the limit where the PDC pump beam can be approximated by a monochromatic plane-wave of frequency $\omega_0$ propagating 
along the $z$-axis direction, a condition which is roughly satisfied in our experiment.
The detailed treatment can be found in \cite{brambilla2012}.

For the type I BBO used in our experiment, the pump is extraordinarily polarized while the down-converted field is ordinarily polarized (e-oo phase-matching), so that the refractive index $n_0(\theta_0^{\rm SFG},\lambda)$ of the pump depends on its direction angle with respect to the crystal axis $\theta_0^{\rm SFG}$, while the PDC field refractive index $n_1(\lambda)$ is isotropic.
We take into account the possibility that the two crystals are not perfectly tuned, i.e. we consider
conditions where the orientation angles of the two crystals with respect to the pump axis,
denoted by $\theta_0^{\rm PDC}$ and $\theta_0^{\rm SFG}$ respectively, are slightly different.
The reference $k$-vector of the extraordinary wave $k_0^{\rm cr}=k_0(\theta^{\rm cr},\Omega=0)$
can thus differ in the first (cr=PDC) and in the second crystal (cr=SFG).
\begin{figure}[H]
\centering
{\scalebox{0.7}{\includegraphics*{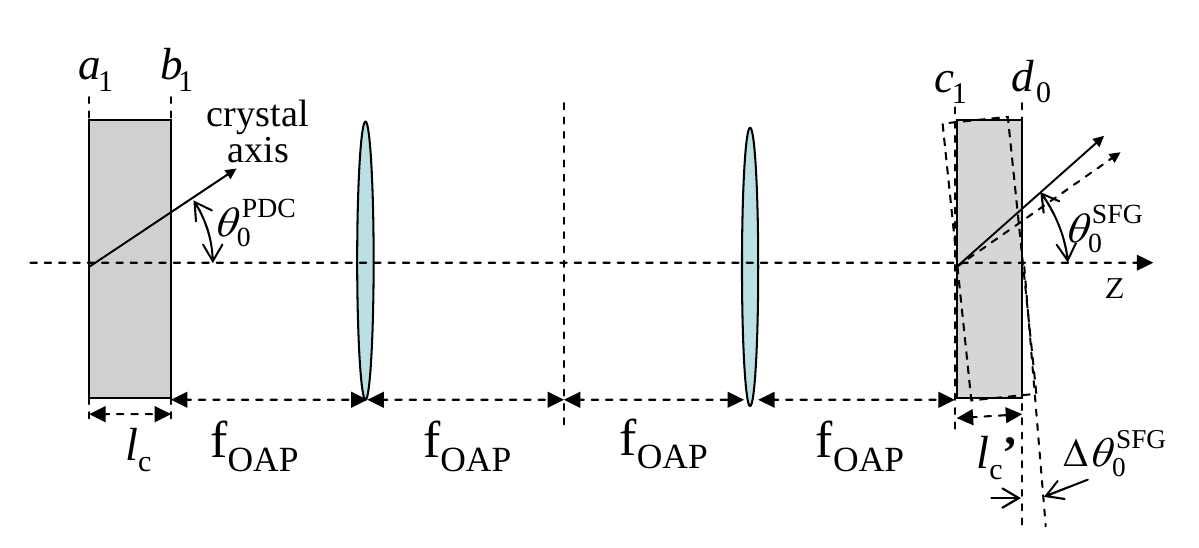}}}
\caption{(color online) Unfolded imaging scheme corresponding to the experimental setup shown in Fig.\ref{fig_setup}. The PDC/SFG field operators
that transform according to the input-output relations (\ref{inputoutput1}) and (\ref{inputoutput_sfg}) and the crystal orientation angles with respect
to the pump axis are indicated in the figure. The rotation $\Delta\theta_0^{SFG}=\theta_0^{SFG}-\theta_0^{PDC}$ applied to the second crystal with respect to the first
is exaggerated in the figure.}
\label{fig_scheme}
\end{figure}

For briefness we shall use the shorthand notation $\vka\equiv(\q,\Omega)$ for denoting a Fourier mode of temporal frequency $\omega_j+\Omega$ and transverse wave-vector $\q$, with the reference frequencies $\omega_j$ denoting either the pump frequency $\omega_0$ when dealing with the extraordinarily polarized up-converted field,
or the degenerate frequency $\omega_1=\omega_0/2$ when dealing with the ordinarily polarized PDC field.
Within the plane-wave pump approximation (PWPA),
the input-output transformation for the PDC field operators
is best written in the Fourier domain where it takes the simple form
\beq
b_1( \vka )=U(\vka)a_1(\vka)+V(\vka)a_1^\dagger(-\vka)\;,
\label{inputoutput1}
\eeq	
where $b_1(\vka)$ and $a_1(\vka)$ are the Fourier transform of the field operator envelopes (carrier frequency $\omega_1=\omega_0/2)$
at the crystal input and output face respectively.
The explicit expression of the gain functions $U(\vka)$ and $V(\vka)$ can be found e.g. in \cite{kolobov99,brambilla2004}.
Here we notice that the dependence on the spatio-temporal frequencies $\vka=(\q,\Omega)$ occurs only through the PWP phase-matching function
\beq
\label{deltaPW}
\DeltaPDC(\vka) = k_{1z}(\vka)+k_{1z}(-\vka) -k_0^{\rm PDC}
\eeq
where $k_0^{\rm PDC}=\frac{\omega_0}{c}n_0(\theta_0^{\rm PDC},\Omega=0)$ denotes the wave-number of the extraordinarily polarized plane-wave pump, $k_{1z}(\vka)=\sqrt{\frac{\omega_1+\Omega}{c}n_1(\omega)-q^2}$ is the wavevector longitudinal component for the ordinary PDC mode $\vka$.
All the properties of the PDC light are described by the following second-order field correlation functions
\bsub
\label{corr_PDC}
\beqa
&&\langle b_1^{\dagger}(\vka)b_1(\vkap)\rangle=\delta(\vka-\vkap)|V(\vka)|^2\label{corr_PDCa}\\
&&\langle b_1(\vka)b_1(\vkap)\rangle=\delta(\vka+\vkap)  U(\vka)V(-\vka)\label{corr_PDCb}\;,
\eeqa
\esub
In particular, the PDC spatio-temporal spectrum
\bsub
\label{PDC_spectrum}
\beqa
&&{\cal S}_{\rm PDC}(\vka)\equiv|V(\vka)|^2= g^2\frac{\sinh^2 [\Gamma(\vka)l_c] }{\Gamma^2(\vka)l_c^2}\\
&&\Gamma(\vka)l_c
      =\sqrt{g^2-\frac{[\DeltaPDC(\vka)l_c]^2}{4}}\;,
      \label{Gammalc}
\label{V}
\eeqa
\esub
gives the PDC photon number distribution in the spatio-temporal frequency domain. The dimensionless parametric gain $g$, proportional to the pump amplitude, the PDC crystal length $l_c$ and the effective $\chi^{(2)}$ coefficient, determines the number of photons $\sinh^2 g$
for perfectly phase-matched mode, for which $\Delta(\vka)=0$.

The other quantity of interest is the spectral probability amplitude (also called biphoton amplitude)
of down-converting a pair of photons in the phase-conjugated modes $\vka$ and$-\vka$
\beqa
\label{UV}
&&\UV(\vka)\equiv U(\vka)V(-\vka)
 =g e^{ik_0 l_c}\frac{\sinh\Gamma(\vka)l_c}{\Gamma(\vka)l_c}
\left\{\phantom{\frac{a}{b}}\hspace{-.3cm} \cosh\Gamma(\vka)l_c+i\frac{\DeltaPDC(\vka)}{2\Gamma(\vka)}
     \sinh\Gamma(\vka)l_c \right\}\,.
\eeqa

For the PDC gain considered in the experiment, only a small fraction of the injected PDC light is up-converted in the second crystal through SFG,
and the propagation equation describing the SFG process can be solved accordingly within a perturbative approach.
We obtain in this way the following expression linking the operators of the SFG field at the crystal output plane,
$d_0(\vka)$, to those of the PDC field imaged at the input plane, $c_1(\vka)$:
\begin{widetext}
\beqa
d_0(\vka)
= e^{\im k_{0z}(\vka) l_c'}   \left[
c_0(\vka)
-
\int\frac{d\vkap}{(2\pi)^{3/2}}
         c_1(\vka-\vkap)
         c_1(\vkap)
		\Si (\vka-\vkap,\vkap)        \right] \, ,
\label{inputoutput_sfg}
\eeqa
\end{widetext}
where $l_c'$ denotes the SFG crystal length, the input envelope operator $c_0$ at the career frequency $\omega_0$ is taken in the vacuum state
(it is assumed that the PDC pump field is completely subtracted by the glass filter shown in Fig.\ref{fig_setup}), and
\bsub
\beqa
\label{Sidef}
&&\hspace{-1.cm} \Si (\vka,\vkap)=\sigma l_c'
e^{i\frac{\Delta(\vka, \vkap)l_c'}{2}}
                        \sinc\frac{\Delta(\vka, \vkap)l_c'}{2} \, ,\\
&&\hspace{-1.cm}\Delta(\vka, \vkap)l_c'=[
k_{1z}(\vka)+k_{1z}(\vkap)
-k_{0z}(\vka+\vkap)]l_c',\label{deltaSFG}
\eeqa
\esub
is the probability amplitude density describing the SFG process:
its square modulus is proportional to the probability for a pair of photons in the fundamental mode $\vka\equiv(\q,\Omega)$
and $\vka'\equiv(\q',\Omega')$ to be up-converted into the second-harmonic mode $\vka+\vka'\equiv(\q+\q',\Omega+\Omega')$.
This up-conversion probability is non negligible only for those mode pairs for which the phase-mismatch $\Delta(\vka, \vkap)l_c'$ accumulated in the SFG crystal
is sufficiently small.
Both the coupling constant $\sigma$ in Eq.(\ref{Sidef}), and the PDC parametric gain $g$ are proportional to the effective $\chi^{(2)}$ coefficient of the crystal for the chosen phase-matching conditions \cite{brambilla2012}.

In Ref.\cite{brambilla2012} we show that the power spectrum $\langle d_0^{\dagger}(\vka)d_0(\vka)\rangle$
of the up-converted SFG field evaluated at the output
face of the SFG crystal can be written as the sum of a coherent and an incoherent component, the former being given by
\beqa
&&\langle d_0^{\dagger}(\vka)d_0(\vka)\rangle_{coh}\nn\\
&&=
\left|\delta(\vka)
\int\frac{d\vkap}{(2\pi)^{3/2}}
\UV(\vkap)
\Si(\vkap;-\vkap)
\right|^2,\label{Scoh}
\eeqa
while the incoherent component is defined through the relation
\bsub
\label{S_PWPA}
\beqa
&&\langle d_0^{\dagger}(\vka)d_0(\vkap)\rangle_{inc}
=\delta(\vka-\vkap){\cal S}_{\rm SFG}^{\rm inc}(\vka)\\
&&{\cal S}_{\rm SFG}^{(\rm inc)}(\vka)
=2
 \int\frac{d\vkap}{(2\pi)^{3}}
 {\cal S}_{\rm PDC}(\vka-\vkap){\cal S}_{\rm PDC}(\vkap)
 |\Si(\vka-\vkap;\vkap)|^2.\label{Sincoh}
\eeqa
\esub

The coherent SFG component (\ref{Scoh}) originates from the up-conversion of
pairs of phase-conjugate PDC modes, $\vka$ and $-\vka$,
into the original plane-wave pump mode $\vka=0$,
that leads to the partial reconstruction of the original coherent pump beam \cite{jedr2011,dayan2007}.
The efficiency of this coherent process is indeed determined by the PWP phase-matching function inside the the SFG crystal, i.e.
\beq
\DeltaSFG(\vka)\equiv\Delta(\vka,-\vka)=k_{1z}(\vka)+k_{1z}(\vka)-k_0^{\rm SFG}\label{deltaSFG_coh}
\eeq
It has been demonstrated in previous works, both theoretically \cite{brambilla2012} and experimentally \cite{jedr2012a,jedr2012b}, that the whole X-shaped structure of the PDC correlation function in the space-time domain can be retrieved by monitoring the coherent SFG component while performing
a careful manipulation of the spatial and the temporal degrees of freedom of the PDC field between the two $\chi^{(2)}$ crystals.

We shall now focus on the incoherent component of the spectrum (\ref{Sincoh}), the main target of our investigation, which
derives from the up-conversion of pairs of photons which do not belong to phase-conjugate modes
and give rise to the speckle-like background observed in the experiments reported in Refs.\cite{jedr2011,jedr2012a,jedr2012b}.
As it can be inferred from Eq.(\ref{Sincoh}), it reduces to the self-convolution of the PDC spectrum in the limiting case where the SFG crystal is extremely thin, since the SFG probability amplitude $\Si$ can be replaced by unity under the integral sign in Eq.(\ref{Sincoh})
when $l_c'\ll l_c$.
We shall verify, however, that propagation effects inside the SFG crystal become relevant as soon as realistic lengths of the SFG crystal are considered, giving the SFG incoherent spectrum a peculiar skewed geometry in the spatio-temporal frequency domain, thus highlighting
the role of interplay between space and time.

\section{Evaluation of the incoherent SFG spectrum}
More insight about the structure of the incoherent spectral component (\ref{S_PWPA}) can be gained
by close inspection of the phase-matching functions in the SFG crystal (\ref{deltaSFG}).
With this goal in mind, we separate the longitudinal $k$-vector component of the ordinary PDC field $k_{1z}(\vka)=\sqrt{k_1^2(\Omega)-q^2}$ into its even and odd parts
\beqa
\label{k1z_decomp}
&&\hspace{-.5cm}k_{1z}(\vka)
=
\frac{k_{1z}(\vka)+k_{1z}(-\vka)}{2}
+
\frac{k_{1z}(\vka)-k_{1z}(-\vka)}{2}
\eeqa
We notice that the even part can be expressed in terms of the PWP phase-matching function (\ref{deltaPW})
\beq
\frac{k_{1z}(\vka)+k_{1z}(-\vka)}{2}
=
\frac{\DeltaPDC(\vka)}{2}+\frac{k_0^{\rm PDC}}{2}
\label{k1z_even}
\eeq
while the odd part can be expanded in odd powers of $q$ and $\Omega$
\beq
\frac{k_{1z}(\vka)-k_{1z}(-\vka)}{2}
=
k_1'\Omega+\frac{1}{6}k_1'''\Omega^3+\frac{k_1'}{2k_1^2}q^2\Omega\label{k1z_odd}
+\hdots
\eeq
where we used the shorthand notations $k_1\equiv k_1(\Omega=0)$,
$k_1' = d k_1/ d\Omega |_{\Omega=0}$,  $k_1'' = d^{2} k_1/ d\Omega^{2} |_{\Omega=0}$, etc. 
and we took into account the independence of the PDC dispersion relation $k_1(\Omega)=\frac{\omega_1+\Omega}{c}n_1(\Omega)$ on
the propagation direction (i.e. on the transverse wave-vector $\q$).
By neglecting the cubic terms in Eq.(\ref{k1z_odd}) (the third order dispersion and the term $\propto q^2\Omega$), the SFG phase-matching function
that enters into the incoherent spectrum expression (\ref{Sincoh}) can be cast into the form
\beq
\Delta(\vkap,\vka-\vkap)
\approx
{\cal D}^{\rm inc}(\vka)
+
\frac{1}{2}
\left\{
\DeltaPDC(\vkap)+\DeltaPDC(\vka-\vkap)
\right\},
\label{pm_incoh}
\eeq
with
\beq
{\cal D}^{\rm inc}(\vka)=k_0^{\rm PDC}-k_{0z}(\vka)+k_1'\Omega\,.
\label{Dinc}
\eeq
This result is very useful because it allows us to rewrite the incoherent SFG spectrum (\ref{Sincoh})
in the form
\begin{widetext}
\bsub
\label{S2_PWPA}
\beqa
{\cal S}_{\rm SFG}^{(\rm inc)}(\vka)
&\propto&
 \int\frac{d\vkap}{(2\pi)^{3}}
 {\cal S}_{\rm PDC}(\vka-\vkap){\cal S}_{\rm PDC}(\vkap)\\
&&\;\times 
\sinc^2\frac{1}{2}\left\{{\cal D}^{\rm inc}(\vka)l_c'+\frac{l_c'}{2l_c}
\left[
\DeltaPDC(\vkap)l_c+\DeltaPDC(\vka-\vkap)l_c
\right]
\right\}
 \label{Sincoh2}\\
&\approx& 
\sinc^2\left[\frac{{\cal D}^{\rm inc}(\vka)l_c'}{2}\right]
\int\frac{d\vkap}{(2\pi)^{3}}
 {\cal S}_{\rm PDC}(\vka-\vkap){\cal S}_{\rm PDC}(\vkap)
\eeqa
\esub
\end{widetext}
The last expression involves a rather rough approximation based on the observation that  
the presence of ${\cal S}_{\rm PDC}(\vka)$ and ${\cal S}_{\rm PDC}(\vka-\vkap)$ under the integral in Eq. (\ref{Sincoh2}) forces
$\DeltaPDC(\vkap)l_c\approx 0$ and $\DeltaPDC(\vka-\vkap)l_c\approx 0$ in the whole integration region,
due to the fact that the PDC spectrum is strongly peaked around the region where phase-matching occurs.
Strictly speaking it should be valid only in the limit $l_c\gg l_c'$, however it turns out to give a good description
of the SFG spectrum also when the two crystals have similar lengths.

\begin{figure*}
\centering
{\scalebox{.88}{\includegraphics*{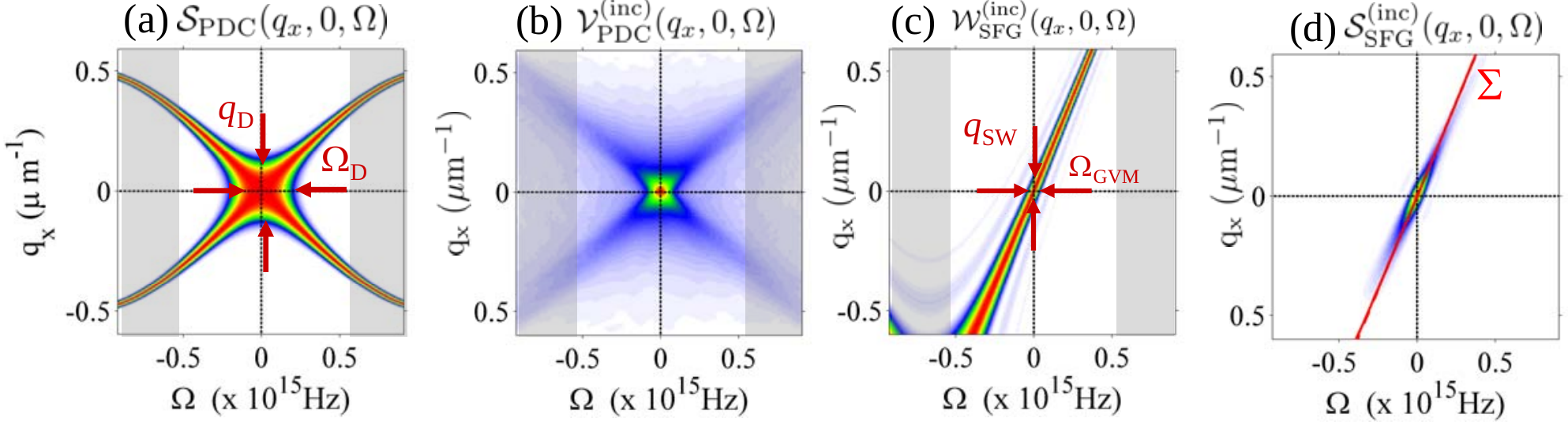}}}
\caption{(color online) Density plots of (a) the PDC spectrum ${\cal S}_{\rm PDC}(\vka)$, the functions (b) $\cV(\vka)$, (c) $\cW(\vka)$ and
the incoherent SFG spectrum ${\cal S}_{\rm SFG}(\vka)$in the $(q_x,\Omega)$-plane. The PDC and SFG crystal lengths are respectively $l_c=4000\,\mu$m and $l_c'=1000\,\mu$m, the PDC parametric gain $g=9.3$, $\Delta_0^{\rm PDC}=\Delta_0^{\rm SFG}=0$. The dashed gray zone indicates the box-shaped frequency filter used in the simulation
to select a temporal PDC bandwidth of $\sim 10^{15}$Hz.
($l_c=4000\,\mu$m, $l_c'=1000\,\mu$m, $g=9.3$. $\theta_0^{\rm SFG}=\theta_0^{\rm SFG}=23^{\circ}$)}
\label{fig_inc}
\end{figure*}

Assuming the validity of this approximation, the SFG spectrum takes thus the factorized form
\beqa
\label{calS2}
&&
{\cal S}_{\rm SFG}^{(\rm inc)}(\vka)
\approx \cW(\vka)\cV(\vka).
\eeqa
where the first factor
\beqa
&&\cW(\vka)=(\sigma l_c')^2\sinc^2 \frac{{\cal D}^{\rm inc}(\vka)l_c'}{2},\label{cW}
\eeqa
describes the effect of propagation in the SFG crystal and determines the overall shape of spectrum.
The second factor, which does not depend on the SFG crystal parameters,
\beqa
&&\cV(\vka)=
\int\frac{d\vka}{(2\pi)^{3}}
 {\cal S}_{\rm PDC}(\vka-\vkap){\cal S}_{\rm PDC}(\vkap)\;\;\;\;\;
 \label{cV}\,
\eeqa
is the self-convolution of the PDC spectrum given by Eq.(\ref{PDC_spectrum}) and acts only as a slow modulation. 
This can be clearly seen in the numerical example of Fig.\ref{fig_inc}, which shows the functions in play in the $(q_x,\Omega)$ plane
(along the walk-off direction)
in the case of two type I BBO crystals aligned for collinear phase-matching at degeneracy.
The first plot shows the PDC spectrum with its characteristic bandwidths $q_D$ and $\Omega_D$ (see the discussion at the end of this section).
The second plot is the self-convolution of such a spectrum, i.e. the function $\cV(\vka)$ appearing in the SFG spectrum in Eq.(\ref{calS2}). 
The third plot displays $\cW(\vka)$ with the corresponding bandwidth $q_{SW}$ and $\Omega_{GVM}$.
The last plot shows the SFG spectrum evaluated from the full expression (\ref{Sincoh}), without using the factorized approximation (\ref{calS2}).
It indeed demonstrates that the SFG spectrum is roughly the product of the functions $\cV$ and $\cW$, in agreement with
Eqs.(\ref{calS2})-(\ref{cV}), and that the shape of the spectrum is mainly determined by the function $\cW(\vka)$.

We thus see that the spectrum in the $(q_x\Omega)$-plane has a skewed cigar-like shape. In the full 3D Fourier domain this corresponds to a volume
centered around a skewed surface $\Sigma$ determined by the region in $(\q,\Omega)$-space where $\cW(\vka)$ takes its maximum value, i.e.
\beq
\label{Sigma}
\Sigma:
{\cal D}^{\rm inc}(\q,\Omega)=0.
\eeq
A closer insight can be gained by exploiting
the following approximation for the longitudinal $k$-vector component of the ordinary and extraordinary fields
\bsub
\label{k01z_quad}
\beqa
&&k_{1z}(\q,\Omega)
\approx k_1+k_1'\Omega+\frac{1}{2}k_1''\Omega^2-\frac{q^2}{2k_1}\;,\label{k1z_quad}\\
&&k_{0z}(\q,\Omega)
\approx k_0^{\rm SFG}+k_0\,'\Omega+\frac{1}{2}k_0\,''\Omega^2-\rho_0\, q_x-\frac{q^2}{2k_0^{\rm PDC}},
\label{k0z_quad}
\eeqa
\esub
where the walk-off angle $\rho_0=-\partial k_0/\partial q_x$ and the $k$-vector derivative $k_0'$ and $k_0''$ of the extraordinary field
are meant to be evaluated at $\Omega=0$ along the optical axis $z$ (the $x$-axis is taken along the walk-off direction).
By substituting Eq.(\ref{k0z_quad}) into Eq.(\ref{Dinc}) we obtain the approximated expression
\begin{widetext}
\beqa
\label{Dincoh}
{\cal D}^{\rm inc}(\q,\Omega)l_c'\approx
(k_0^{\rm PDC}-k_0^{\rm SFG})l_c'
+\frac{q_x}{q_{\rm SW}}-\frac{\Omega}{\Omega_{\rm GVM}}
+\frac{l_c'}{2k_0^{\rm SFG}}q^2-\frac{k_0''l_c'}{2}\Omega^2.
\label{Sincoh3b}
\eeqa
\end{widetext}
where
\beqa
&&\Omega_{\rm GVM}
=
\frac{1}{(k_0'-k_1')l_c'},\;\;\;\;
\label{omg_GVM}
q_{\rm SW}
=
\frac{1}{\rho_0 l_c'},\;\;\;
\label{q_SW}
\eeqa
determines the widths of $\cW(\q,\Omega)$ in the spatial and temporal frequency domains respectively, as indicated in
the example of Fig.\ref{fig_inc}c. For small $q$ and $\Omega$
the linear terms in Eq.(\ref{Dincoh}) are the dominant ones: they describe the effect of GVM and spatial walk-off between the PDC and SFG
fields in the incoherent up-conversion process,
$q_{\rm SW}^{-1}$ and $\Omega_{\rm GVM}^{-1}$
giving respectively the typical transverse separation and the time delay between the ordinary (PDC) and the extraordinary (SFG) fields
after a propagation distance equal to the SFG crystal length $l_c'$.

Neglecting the quadratic terms with respect to the linear ones,
the surface $\Sigma$ is close to the skewed plane 
\beq
\label{Sigma_prime}
\Sigma':
\frac{q_x}{q_{\rm SW}}=
\frac{\Omega}{\Omega_{\rm GVM}}-
(k_0^{\rm PDC}-k_0^{\rm SFG})l_c'
\eeq
Notice that the constant term, as we shall see in the next section, is non vanishing only when the two crystals are tuned
for different phase-matching conditions.
We see from relation (\ref{Sigma_prime}) that incoherent upconversion from PDC occurs only for those spatio-temporal modes 
for which the GVM between the fundamental and the up-converted light is compensated by their spatial walk-off.

We conclude this section by observing that, 
using the same quadratic approximation (\ref{k01z_quad}), the
phase-matching function in the PDC crystal can be written as
\beq
\DeltaPDC(\q,\Omega)l_c=(2k_1-k_0^{\rm PDC})l_c+\frac{\Omega^2}{\Omega_D^2}-\frac{q^2}{q_D^2},
\eeq
where the introduced bandwidths associated with GVD and diffraction
\beq
\Omega_D=\sqrt{1/k_1''l_c},\;\;\;\;
q_D=\sqrt{k_1/l_c},
\label{PDC_band}
\eeq
give an estimate of the spectral widths of both ${\cal S}_{\rm PDC}(\vka)$
and $\cV(\vka)$ along the temporal and the spatial frequency axes respectively
(as indicated in the plot of Fig.\ref{fig_inc}a).
Unless the SFG crystal is extremely thin, the typical variation scale ($q_{SW}$, $\Omega_{GVM})$ of $\cW(\vka)$ are much 
shorter than ($q_D$, $\Omega_D)$ (see Fig.\ref{fig_inc}a,c). This is the reason why
the self-convolution of the PDC spectrum (\ref{cV}) acts as a slow modulation, on the scales ($q_D$, $\Omega_D)$,
while $\cV(\q,\Omega)$ defines the overall shape of the incoherent spectrum.
More precisely,  we find that
$\cW(\vka)$ becomes narrower than  $\cV(\vka)$
for
\beq
\label{lcp_SW}
q_{SW}<q_D\longleftrightarrow l_c'>\sqrt{\frac{l_c}{k_0}} \rho_0\approx 150\,{\rm \mu m}
\eeq
in the spatial frequency domain, and for
\beq
\label{lcp_GVM}
\Omega_{GVM}<\Omega_D\longleftrightarrow l_c'>\frac{\sqrt{k_0''l_c}}{k_0'-k_1'}\approx \,360 {\rm \mu m}
\eeq
in the temporal frequency domain, the given numerical values being estimated for the 4mm type I BBO crystal used in our experimental setup.
For $l_c'$ exceeding those values, spatial walk-off and GVM becomes the dominant
processes affecting the SFG incoherent emission, as the ratio $q_{SW}/q_D$ and $\Omega_{GVM}/\Omega_D$
decrease below unity scaling as $1/\sqrt{l_c'}$.

\section{Incoherent spectrum vs SFG crystal length and angular tuning}
\label{sec:spectrum_num}
In this section we give an overview of the incoherent spectrum behaviour with respect to different
phase-matching conditions in the SFG crystal, obtained by changing either its length or its relative orientation
with respect to the PDC crystal.
The numerical results are obtained with a full 3D+1 modeling
of the experiment \cite{jedr2011,jedr2012a,jedr2012b} in order to achieve a realistic description
of the optical system.

The generation of the broadband PDC field is described by stochastic simulations in the framework of the
Wigner representation (see e.g. \cite{brambilla2004,brambilla2012}), which take into
account both the finite cross section and duration of the pump pulse and the phase-matching conditions inside
both crystals (the Sellmeier dispersion relation for the BBO crystal found in \cite{boeuf2000} are used).
The waist and the duration of the injected Gaussian  pump pulse are respectively $w_p=500\mu$m and $\tau_p=1$ps, the PDC gain parameter
$g\approx 9$ corresponds to a pulse energy close to $350\,{\rm \mu J}$.
We verified that the PWPA model provides similar results for the chosen pump parameters.

\begin{figure*}
\centering
{\scalebox{.93}{\includegraphics*{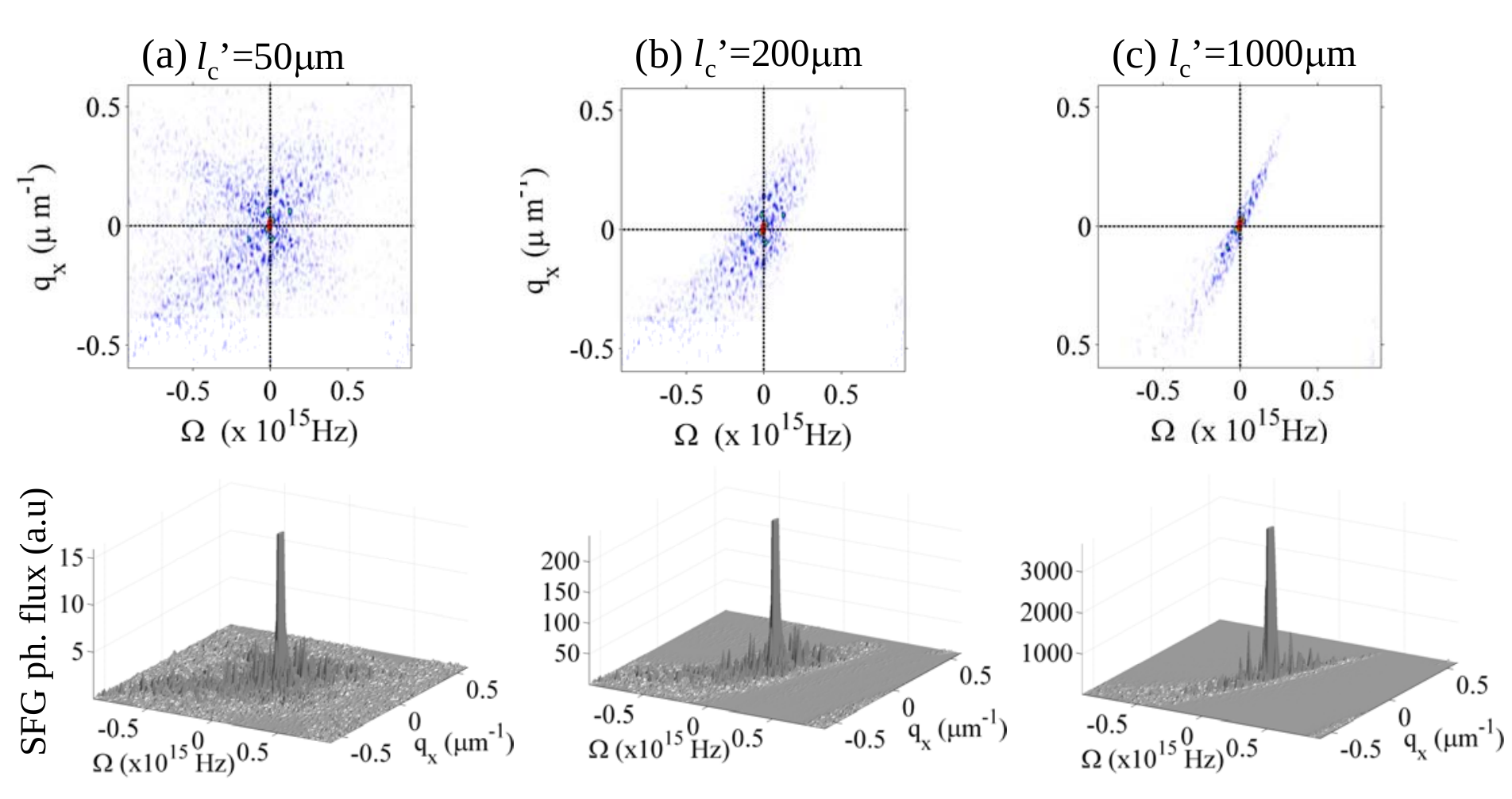}}}
\caption{(color online) SFG spectrum in the $(q_x,\Omega)$-plane as obtained from the 3D+1 model
for same parameters as in Fig.\ref{fig_inc}.
The central peak at $(q_x=0,\Omega=0)$ corresponds to the coherent component of the spectrum. The speckled incoherent
component from a single pump pulse displays a distribution similar to that obtained with
the PWPA model in Fig.\ref{fig_inc}.
Both crystal are oriented for collinear phase-matching at degeneracy, $\tau_p=1$ps, $w_p=500{\rm \mu}$m, $g=9.3$,
$l_c=4000{\rm \mu m}$.}
\label{fig_spectr3D}
\end{figure*}

Figure \ref{fig_spectr3D} shows the evolution of the SFG spectrum for increasing lengths of the SFG crystal $l_c'$.
In each single stochastic realization the incoherent component of the spectrum 
appears as a speckle-like pattern \footnote{Ensemble averages performed by repeating the simulation would provide the correct quantum mechanical mean values for the spectrum in the framework of the Wigner representation as
described e.g. in \cite{brambilla2004}. However, the required CPU time would be prohibitive and would not add more insight to the description.}
(see \cite{brambilla2012} for more details), while the PWPA result (\ref{S_PWPA}) provides the mean value of the spectrum (see e.g. Fig.\ref{fig_inc}d).
When the SFG crystal is sufficiently thin, as in Fig.\ref{fig_spectr3D}a, conditions (\ref{lcp_SW}) and (\ref{lcp_GVM}) are not satisfied and
both spatial walk-off and GVM are almost ineffective. In this limit
the SFG incoherent spectrum reproduces the shape of the PDC spectrum self-convolution (\ref{cV}) in agreement with the PWPA model
result presented in the previous section (this can also be inferred by comparing Fig.\ref{fig_spectr3D}a and fig.\ref{fig_inc}b).
By increasing $l_c'$, we find the predicted transition to the much narrower
cigar-shaped spectrum skewed in the walk-off plane, close to the PWPA
result shown in Fig.\ref{fig_inc}d [its orientation in the $(q_x,\Omega)$-plane
is given by Eq.(\ref{Sigma_prime})].

Notice also that the 3D+1 model includes the coherent component of the SFG spectrum:
the narrow peak centered at $(\q=0,\Omega=0)$, clearly visible in the surface plots  (bottom panel),
corresponds indeed to the coherent reconstruction of the original pump pulse predicted by the PWPA theory [see Eq.\eqref{Scoh}].
For a narrow band pump field spectrum, the coherent peak height is more than four order of magnitude larger
than the widely spread incoherent component.
For this reason the plots in Fig.\ref{fig_spectr3D} have been truncated to $0.05\%$ of the coherent peak value
in order to display the incoherent component distribution. An analytical estimation of the visibility of the coherent SFG component
against the incoherent contribution as a function of the pump beam parameters and the PDC gain will be given elsewhere \cite{nw}.

\begin{figure*}   
\centering
{\scalebox{.75}{\includegraphics*{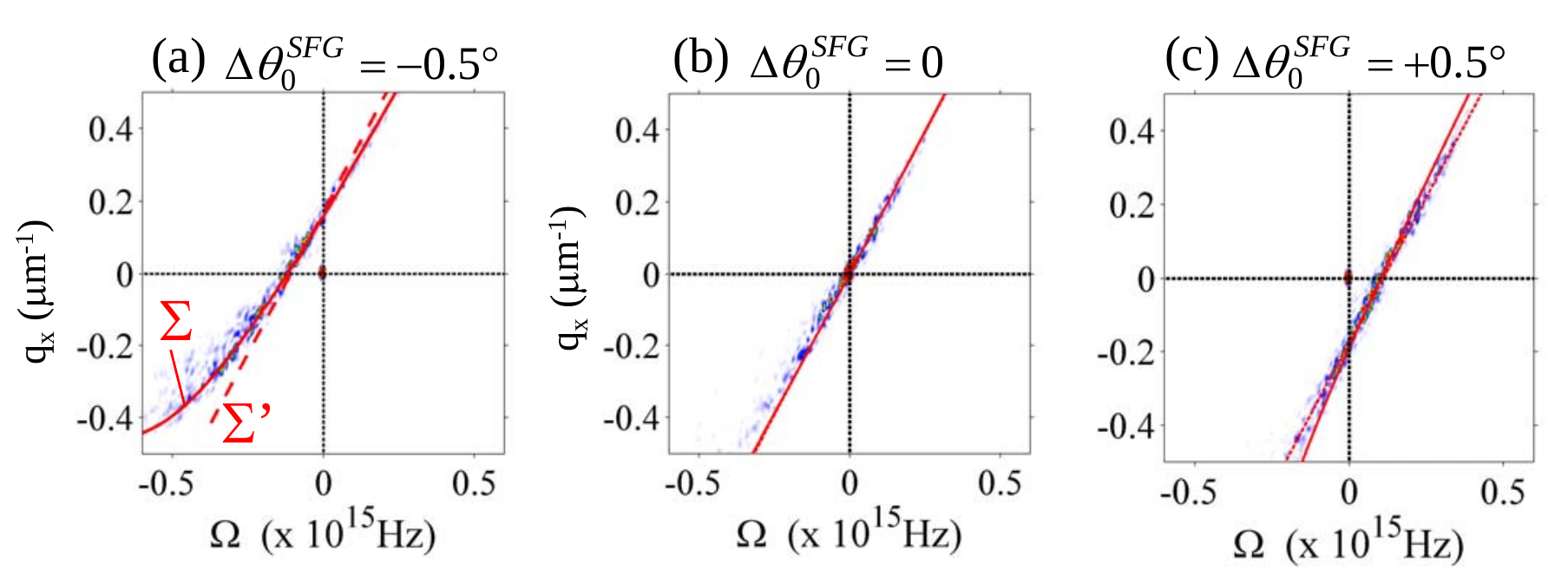}}}
\caption{(color online) SFG spectrum in the $(q_x,\Omega)$-plane (parallel to the walk-off plane) for different mistuning of the SFG crystal $\Delta\theta_0^{\rm SFG}$
obtained with the 3D+1 model. The simulation parameters are $l_c'=l_c=4000\mu$m, $g=9.3$, $w_p=500\,{\rm \mu}$m, $\tau_p=1$ps. The plots of $\Sigma$ (solid red line) and $\Sigma'$ (dashed red line) are evaluated from Eq.(\ref{Sigma}) and Eq.(\ref{Sigma_prime}) respectively.}
\label{fig_spectrY}
\end{figure*}

\begin{figure*}
\centering
{\scalebox{.75}{\includegraphics*{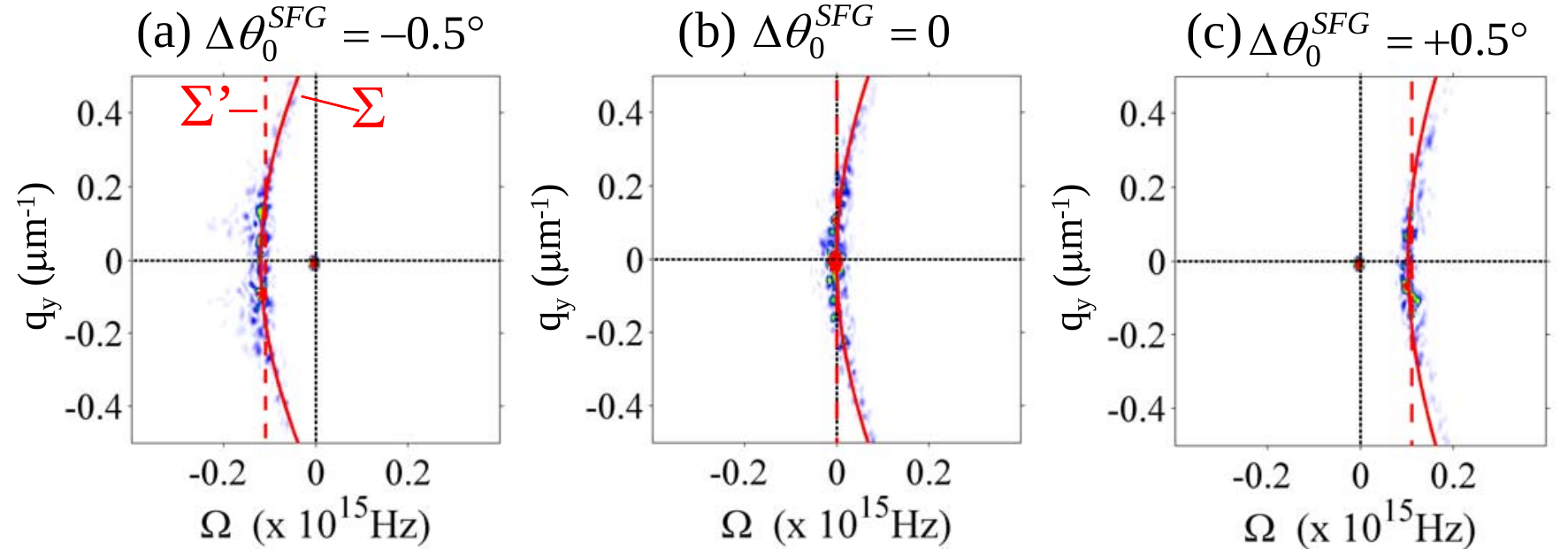}}}
\caption{(color online) SFG spectrum in the $(q_y,\Omega)$-plane (orthogonal to the walk-off plane) for the same parameters in Fig.\ref{fig_spectrY}a}
\label{fig_spectrX}
\end{figure*}

We investigated the effect of a small tilt angle of the SFG crystal with respect to the PDC crystal.
The PDC crystal is oriented for collinear PM as in the previous examples, while the SFG crystal is tilted by the
angles (a) $\Delta\theta_0^{\rm SFG}=-0.5^{\circ}$,(b) $\Delta\theta_0^{\rm SFG}=0^{\circ}$ and (c) $\Delta\theta_0^{\rm SFG}=+0.5^{\circ}$.
Figures \ref{fig_spectrY} and \ref{fig_spectrX} show the SFG spectrum obtained with the 3D+1 model,
evaluated respectively in the plane  parallel and orthogonal to the walk-off direction.
As illustrated in the next section, this corresponds to the two possible configurations of the imaging spectrometer (IS)
implemented in the experiment, the first providing a measure of
${\cal S}_{\rm SFG}(q_y=0,q_x,\Omega)$, the second a measure of ${\cal S}_{\rm SFG}(q_y,q_x=0,\Omega)$.
In the first configuration where ${\cal S}_{\rm SFG}(q_y=0,q_x,\Omega)$ is displayed,
the incoherent spectrum is skewed in the $(q_x,\Omega)$-plane along the
direction direction of $\Sigma'$ (red dashed line in the density plots), as predicted by the PWPA model (\ref{Sigma_prime}).
In the second case, where spatial walk-off does not play any role, the temporal frequencies of the incoherent component
have the same offset $\approx \Delta \Omega_{\rm inc}$ from pump frequency for a large range of $q_y$ spatial frequencies.
This offset is roughly given by
$ \Delta \Omega_{\rm inc}\approx \frac{k_0^{\rm PDC}\rho_0}{k_0'-k_1'}\Delta\theta_0^{\rm SFG}$
as can be inferred by setting $q_x=0$ into Eq.(\ref{Sigma_prime}).
Accordingly, the central wavelength of the incoherent component is
\beqa
\label{Sigmayp}
\lambda_{\rm inc}
\approx
\lambda_0
-\frac{n_0\rho_0\lambda_0}{c(k_0'-k_1')}
\Delta\theta_0^{\rm SFG}
\,.
\eeqa
The red solid line shown in the figures corresponds to the more precise relation between the spatial and temporal frequencies
given by Eq.(\ref{Sigma}),
which includes the slow varying quadratic terms. The latter are responsible of
a slight concavity of the surface $\Sigma$ toward shorter wavelengths, a feature which
fits well the shape of the incoherent spectrum obtained with the 3D+1 model over a broad range of spatial frequencies
(see in particular Fig.\ref{fig_spectrX}a).

\begin{figure}[H]
\centering
{\scalebox{.2}{\includegraphics*{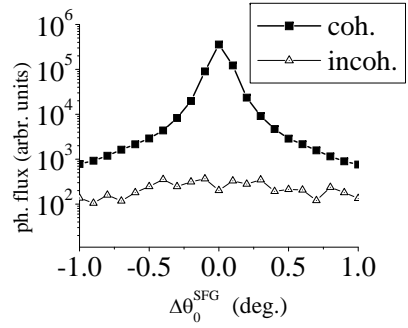}}}
\caption{Number of photons in the coherent and in the incoherent components as a function of $\Delta\theta_c^{\rm SFG}$ as obtained
with the 3D+1 model. The PDC parameters are the same as in the previous figures.}
\label{fig_lshift}
\end{figure}

It is worth noticing that the coherent SFG component rapidly decreases with the mistuning between the two crystals,
while the incoherent peak remains almost unaffected. This behaviour is shown in Fig.\ref{fig_lshift},  where
the number of photons of the two components are plotted as a function of $\Delta\theta_0^{\rm SFG}$.
The coherent up-conversion process, in contrast to the incoherent process, is indeed strongly phase-sensitive and rapidly loose efficiency
when the two crystals are not perfectly tuned. As shown in \cite{brambilla2012}, the number of coherent SFG photons is reduced by about a factor\;10
when the tilt angle $\Delta\theta_c^{\rm SFG}$ exceeds the critical value
\beq
\label{angular_tol}
\Delta\theta_c^{\rm SFG}
=\frac{2\sqrt{\pi^2+g^2}}{\rho_0 k_0 l_c'},
\eeq
which is close to $0.25\,$degree for the chosen BBO parameters ($g=9.3$, $l_c'=4$mm).
Despite that, even for a tilt angle as large as one degree (see Fig.\ref{fig_lshift})
the intensity of the coherent component remains significantly larger than that of the incoherent component 
due to concentration of all the coherent SFG photons in a single mode corresponding to the original pump 
(which is strongly focused in the Fourier domain).

\section{Experimental results}
\label{sec:exp}
The aim of the experimental work was to characterize the spatio-temporal far-field spectrum (in the $(\lambda,\alpha)$ plane) of the SFG radiation, and to verify the 
predicted behaviour of the incoherent component as a function of the angular mistuning between the PDC crystal and the SFG crystal. To this end the output radiation from the SFG crystal was analyzed by means of an imaging spectrometer (grating with $600\,$lines/mm), whose entrance vertical slit (with respect to the optical bench) was placed in the Fourier plane of a $20\,$cm focal length lens. The spectra were recorded in the walk-off plane and in
the plane orthogonal to walk-off.
Because of the geometry of our system and the crystal axis orientation, the vertical plane in the laboratory frame corresponded to the plane orthogonal to walk-off. 
In order to detect the spectra in the walk-off plane, the radiation reconverted from the SFG crystal was tilted by 90 degrees by means of two mirrors aligned in a periscope configuration.
\begin{figure}[H]
\centering
{\scalebox{.5}{\includegraphics*{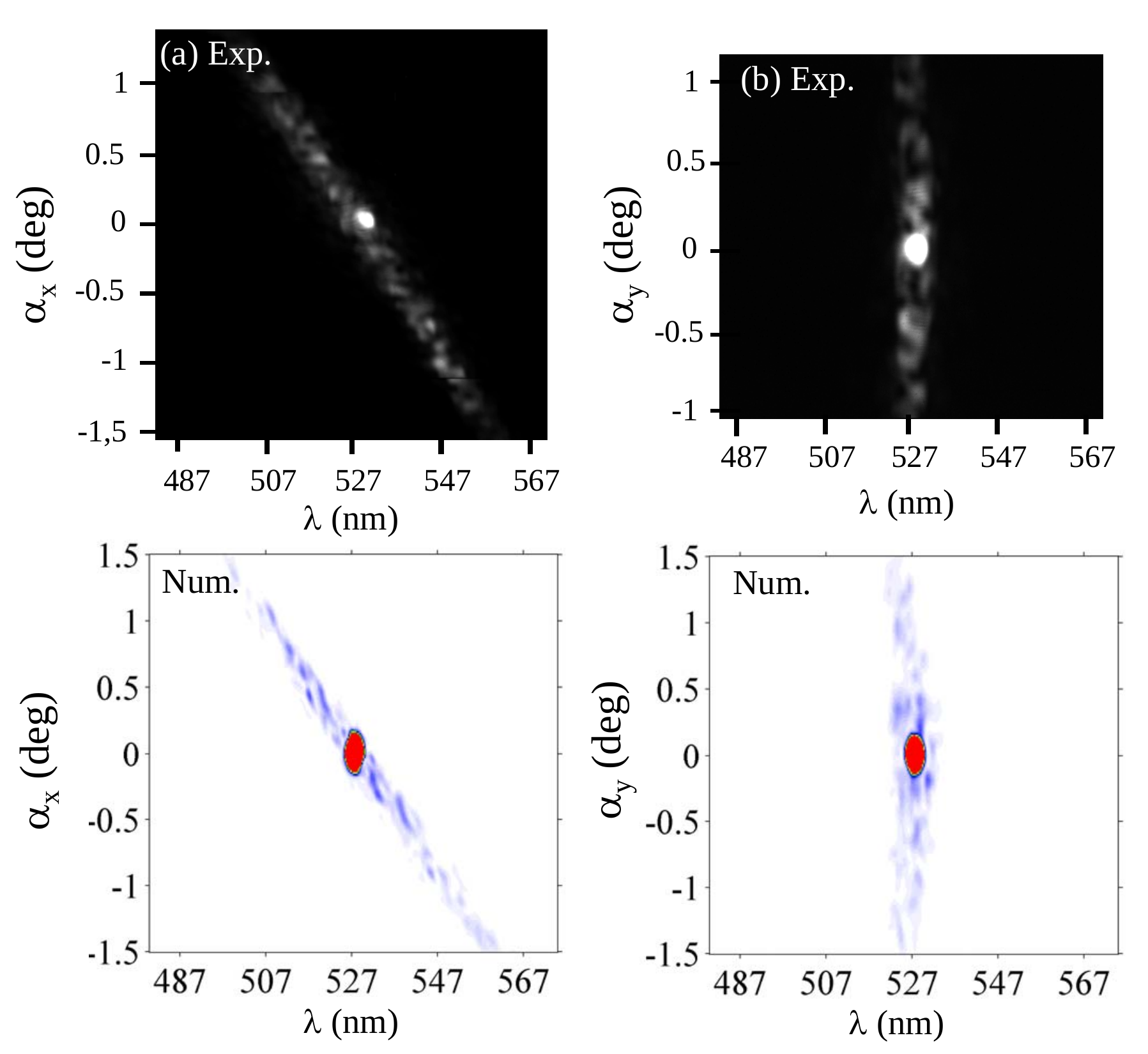}}}
\caption{(color online) Top: SFG far-field spectrum measured from a single pump shot with the IS slit selecting photons emitted (a) in the walk-off direction
and (b) orthogonal to the walk-off direction. The two crystals are perfectly tuned with $\Delta\theta_0^{\rm SFG}=0$.
Bottom: results of the 3D+1 numerical simulation implemented with the same parameters of the experiment.}
\label{fig_1shot}
\end{figure}
The upper panel reported in Fig.\ref{fig_1shot} (gray scale plots) shows typical single shot spectra of the SFG radiation recorded (a) in the plane parallel to walk-off and (b) in the orthogonal plane. The central coherent peak in the collinear direction is evident, and the features of the incoherent component in the $(\lambda,\alpha)$-plane are in agreement with the theory. The bottom panel shows the corresponding spectra obtained with the 3D numerical model. 
The experimental spectra exhibits the skewed geometry predicted by theory, which originates from the compensation of spatial
walk-off and GVM.

\begin{figure}[H]
\centering
{\scalebox{.5}{\includegraphics*{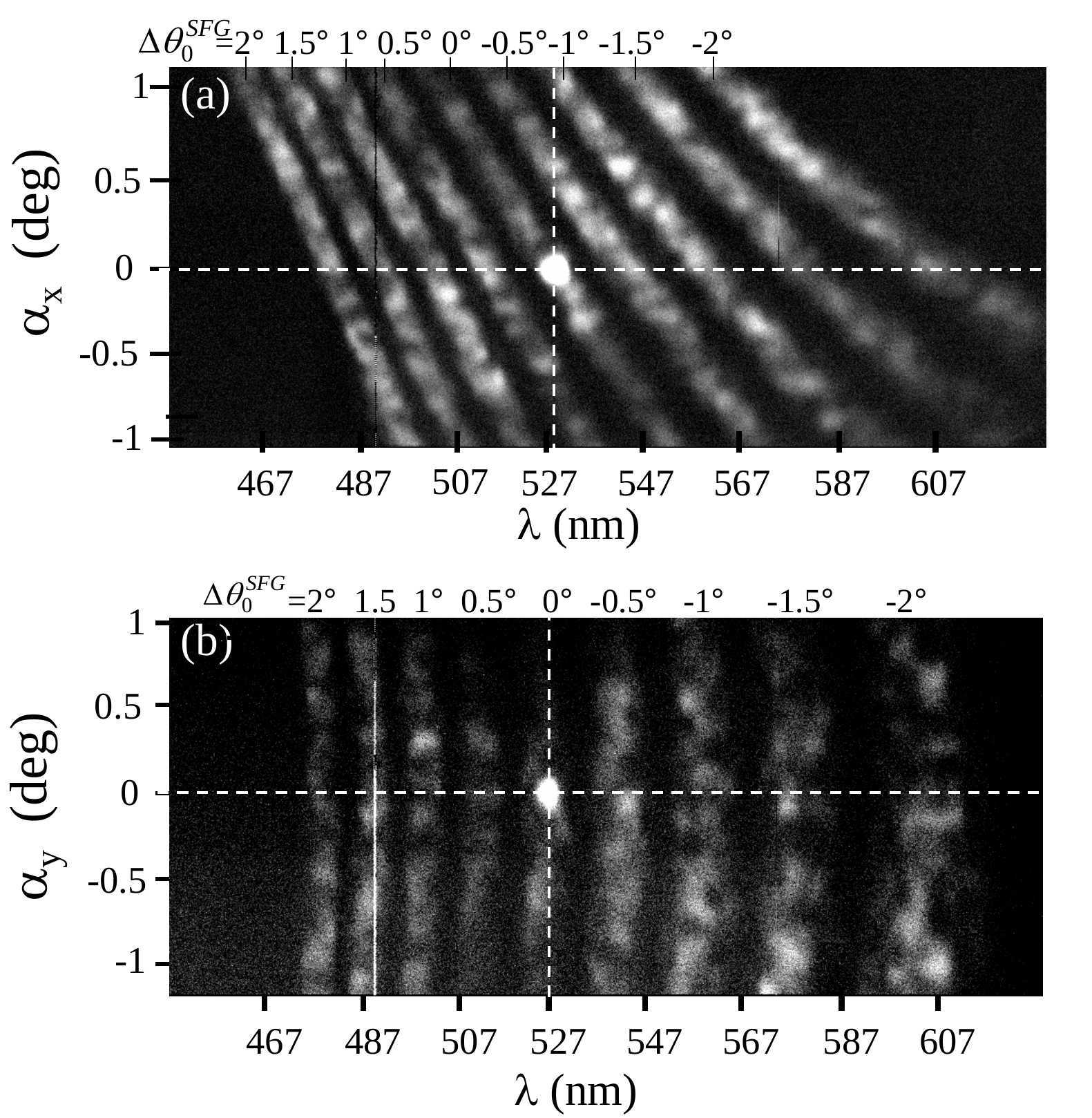}}}
\caption{SFG far-field spectrum measured with the imaging spectrometer for different tuning angles $\Delta\theta_0^{\rm SFG}$ of the SFG crystal (the values are indicated at the figure top). The SFG radiation is recorded by the CCD after integration over 5 laser pump shots for each value of $\Delta\theta_0^{\rm SFG}$.}
\label{fig_XYplane}
\end{figure}

\begin{figure}[H]
\centering
{\scalebox{.3}{\includegraphics*{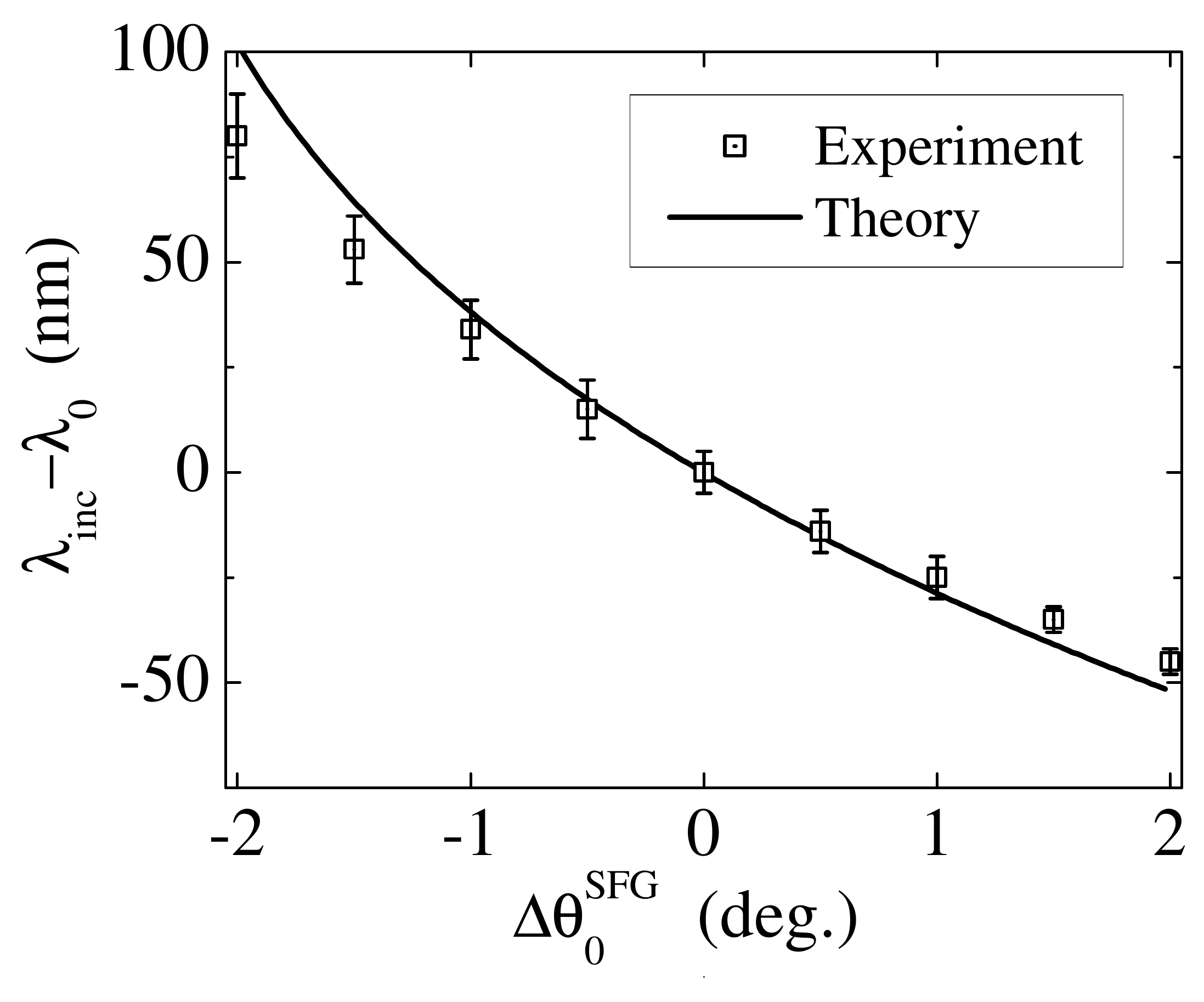}}}
\caption{Plot of the measured central wavelength $\lambda_{\rm inc}$ of the incoherent component along the collinear direction for the applied values of the angular mistuning $\Delta\theta_0^{\rm SFG}$ (square symbols). The solid line is obtained from the PWPA model by solving numerically Eq.(\ref{Sigma}) with respect to $\Omega$ (at $\q=0)$
by varying the value of $\Delta\theta_0^{\rm SFG}$.}
\label{fig_fit}
\end{figure}

Figure \ref{fig_XYplane} illustrates the behaviour of the SFG spectrum when the second crystal is rotated with respect to the first one by an
angle $\Delta\theta_0^{\rm SFG}$; in particular it reports the SFG spectrum in the two different configurations by varying the tuning angle
from $-2^{\circ}$ up to $+2^\circ$ by steps of $0.5^{\circ}$. 
Fig.\ref{fig_fit} displays a more detailed comparison between the theory and the experiment: the square symbols with the error bar reports 
the central wavelength of the incoherent component along the collinear direction ($\q=0$). The solid line in the same figure
is obtained from the PWPA model by solving numerically the equation ${\cal D}^{\rm inc}(\q=0,\Omega)=0$ for different 
tuning angles $\Delta\theta_0^{\rm SFG}$, ${\cal D}^{\rm inc}(\vka)$ being given by Eq.(\ref{Dinc}).
The agreement between theory and experiment is very good although not perfect, especially at large mistuning angles. However,
such a quantitative agreement between theory and experiment is beyond the scope of this work, whose goal is mainly to
demonstrate the characteristic geometry of the spectrum of the SFG light originating from incoherent upconversion of PDC light.

\section{Conclusions}
In this work we investigated the  up-conversion process of broadband PDC radiation, focusing on the spatio-temporal spectral properties 
of its incoherent component.

We have shown that as soon as the SFG crystal is longer than a few hundreds micrometers, the
phase-matching mechanism selects the spatio-temporal frequencies of the incoherent SFG radiation in a non trivial way:   
the spectrum is characterized by a skewed cigar-like shape in the spatio-temporal Fourier
domain which derives from the interplay of group-velocity mismatch and spatial walk-off between the PDC field and the SFG field. 
As for many other nonlinear optical processes, these features are a manifestation of the space-time coupling that occurs because of phase-matching.
We can cite the closely related examples of the skewed coherence along space-time trajectories predicted in three and four-wave mixing processes \cite{picozzi2002}
and of the X-shaped spatio-temporal coherence of twin beams \cite{jedr2006}.  
Other relevant examples are the X-shaped quantum correlation in space-time of twin beams \cite{gatti2009,caspani2010,brambilla2010,jedr2012a,jedr2012b}
and the macroscopic X-waves generated in quadratic media 
\cite{ditrapani2003} or in four-wave mixing process related to a cubic nonlinearity \cite{couairon2006}. 
 
It is only in the limit where the SFG crystal is extremely thin that 
the generated SFG intensity is simply proportional to the square of the injected PDC intensity.
This limit 
corresponds to situations in which phase-matching in the second-crystal does not
play a significant role, so that
the SFG spectrum coincides with the self-convolution of the PDC spectrum.

We have also studied the behaviour of the SFG spectrum with respect to small angular mistuning between the PDC and the SFG crystals.
For increasing values of the angular mistuning, we observe a progressive displacement of the incoherent spectrum with respect
to the wavelength of the coherent peak, a property which could provide a useful tool for optimizing the visibility of twin beam correlation
measurements. 

These results have been illustrated by means of a semi-analytical and a numerical model of the optical system, 
and have been fully confirmed by the experimental measurements.

\acknowledgments
This work was realized in the framework of
the Fet Open project of EC 221906 HIDEAS.

\end{document}